\documentclass[a4paper,11pt]{article}
\usepackage{jinstpub} 
\usepackage{lineno}

\title{\boldmath SPIDER, a Waveform Digitizer ASIC for picosecond timing in LHCb PicoCal}

\author[c]{L. Alvado}
\author[b]{, N. Arveuf}
\author[d]{, E. Bechetoille}
\author[b]{, G. Blanchard}
\author[a,1]{, D. Breton\note{Corresponding author.}}
\author[b,2]{, B. Joly,\note{Speaker and Poster Presenter at TWEPP 2025.}}
\author[c]{, L. Leterrier}
\author[a]{, J. Maalmi}
\author[b]{, S. Manen}
\author[d]{, H. Mathez}
\author[a]{, C. Sylvia}
\author[a]{, P. Vallerand}
\author[b]{, R. Vandaele}

\affiliation[a]{CNRS / IN2P3 / IJCLab, Orsay, France}
\affiliation[b]{CNRS / IN2P3 / LPCA, Clermont-Ferrand, France}
\affiliation[c]{CNRS / IN2P3 / LPC, Caen, France}
\affiliation[d]{CNRS / IN2P3 / IP2I, Lyon, France}

\emailAdd{dominique.breton@ijclab.in2p3.fr}

\abstract{

We present the architecture, design and first test results of SPIDER, the first prototype of a TSMC CMOS 65 nm ASIC designed for the time measurement path of LHCb Electromagnetic Calorimeter after LS4 Upgrade. 

The main requirements for the readout of this detector are a time resolution below 15 ps rms above 5 GeV, and a channel occupancy up to 30\% (12 Mevent/s).

The first prototype called SPIDER\_V0 is a 2-channel waveform digitizer locked on the LHC clock allowing precise time reconstruction by digital algorithms. The architecture is based on 2 DLLs in series controlling respectively the phase of the sampling window  and the sampling frequency, the latter covering the range between 2 and 20 GS/s. Each self-triggering channel houses 8 banks of 32 analog memory cells and a massively parallel Wilkinson ADC for conversion at 5 GHz over 10 bits with a maximal conversion time of 200 ns.

SPIDER targets not only LHCb, but all fast detectors mounted on current and future accelerators. Its sampling frequency can indeed be adjusted to different signal risetimes. Its main frequency of 40 MHz could even be eventually locked to another value by modifying only one of the DLLs in the chip design.}


\begin{document}
\maketitle
\flushbottom

\section{The upgrade of LHCb calorimeter}
\label{sec:calo}

\begin{figure}[htbp]
\centering
\includegraphics[width=1\textwidth]{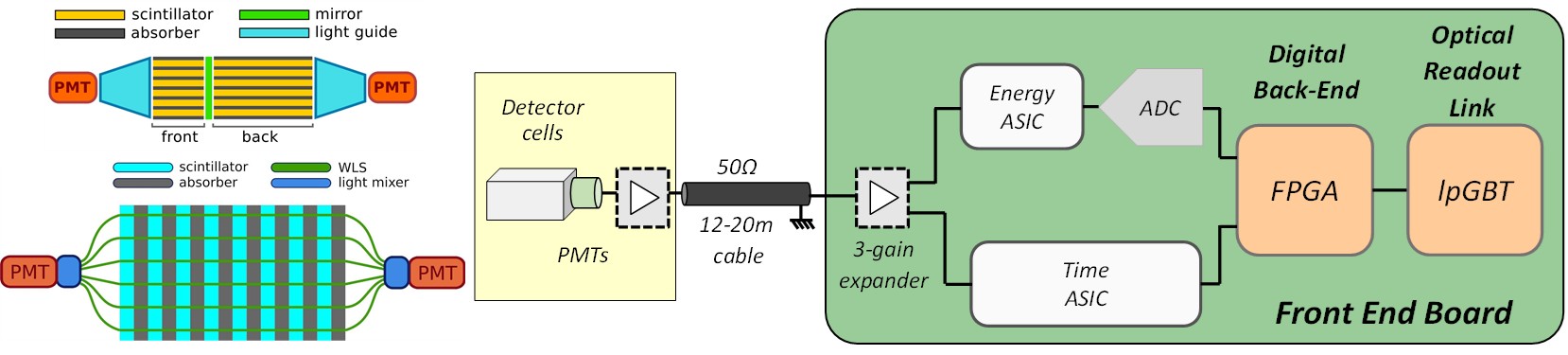}
\caption{Left: SpaCal (top) and Shashlik (bottom) dual readout modules; right: block diagram of PicoCal electronics chain.\label{fig:1}}
\end{figure}

The LHCb electromagnetic calorimeter (ECAL) has to be modified to mitigate radiation damage after Run 3 of LHC and to meet the requirements of Run 5 ~\cite{r1}. To cope with the expected increase in radiation levels and particle densities, particularly in the central region, a significant redesign is foreseen during Long Shutdown 4 (LS4). This upgrade, known as PicoCal, addresses the need for improved radiation hardness, high timing resolution of the order of 15 ps rms above 5 GeV of transverse energy, and granularity ~\cite{r1}. Several scintillating sampling ECAL technologies are currently being investigated in an ongoing R\&D campaign in view of the PicoCal, using Spaghetti Calorimeter (SpaCal) and Shashlik modules. All modules including Shashlik will have double-sided readout, as illustrated in Fig:~\ref{fig:1} (left), with precise timing capabilities.

The performance of both tungsten-based and lead-based prototypes has been evaluated in test beam campaigns covering energy ranges from 1–5 GeV and 20–100 GeV. Beside energy resolution, their timing resolution has also been measured over the full energy ranges, reaching approximately 15 ps at high energies, which is fully compliant with the Upgrade II requirements. Readout electronics will be upgraded in order to introduce time measurement by splitting the chain into two separate ASICs, ICECAL65 (successor of ICECAL~\cite{r2}) developed by the university of Barcelona which will be used for energy measurement, and SPIDER, object of this paper (see Fig:~\ref{fig:1} right). A 3-gain expander will sit in front on the two ASICs in order to provide them signals with adequate amplitudes. All these elements will sit on the front-end boards located on top of the detector.

\section{The SPIDER ASIC}

\begin{figure}[htbp]
\centering
\includegraphics[width=0.49\textwidth]{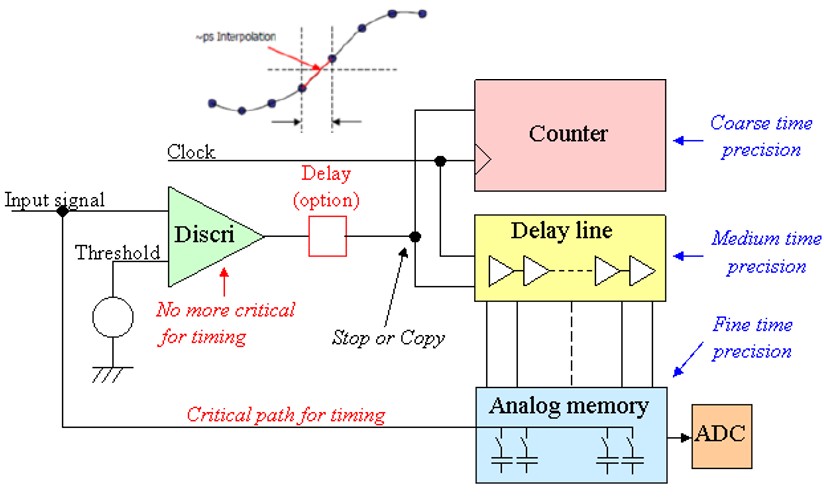}
\includegraphics[width=0.5\textwidth]{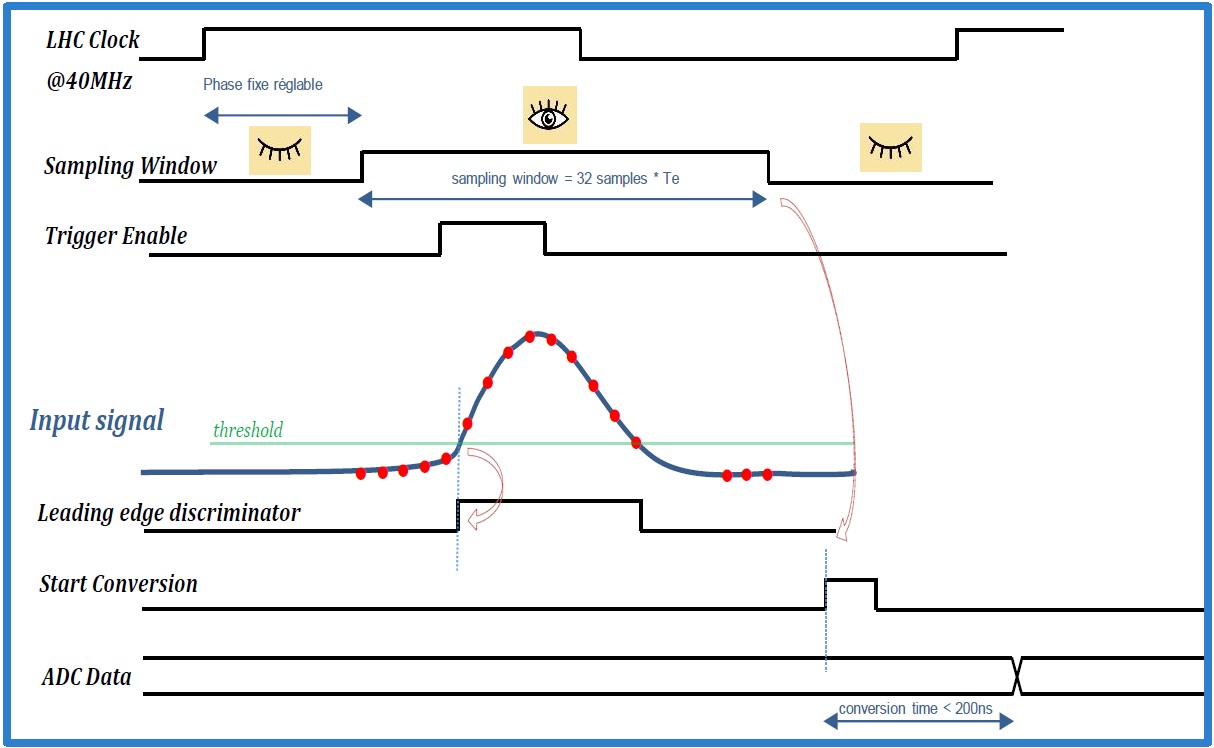}
\caption{Left: principle of Waveform TDC. Right: chronogram of sampling, trigger and conversion in SPIDER.\label{fig:2}}
\end{figure}

The SPIDER ASIC is based on the principle of Waveform Time to Digital Converter (WTDC) formerly introduced by the SAMPIC ASIC~\cite{r3} (see Fig:~\ref{fig:2} left). The latter is a fast digitizer based on analog circular memories. Fine time measurement is performed via interpolation between waveform samples based on the constant fraction discriminator (CFD) algorithm. 

SPIDER is based on the same principle (WTDC) but with a very different architecture. It indeed captures signal within periodic time windows defined in phase relationship with the LHC clock (40 MHz), with adjustable start and stop times fixed by steps of 195 ps (see Fig:~\ref{fig:2}). This selective sampling is allowed by the predictability of signal arrival time in each channel and thus makes a recording depth of only 32 samples sufficient. The targeted signal bandwidth is 2 GHz and the sampling period is configurable between 50 ps and 600 ps, thus sampling window durations of 1.6 ns to 19.2 ns and sampling frequencies of 1.6 to 20 GS/s. 

Each channel integrates a discriminator that can self-trigger independently, and an external trigger is also available. An event occurring during the trigger window causes the current bank to freeze its content at the end of the sampling window, making it available for conversion, and provokes switching to next bank. 
During conversion, all frozen banks are digitized. After conversion, digitized data is extracted via a 10-bit parallel data bus. 
In SPIDER\_V0, the extraction is directly handled by the FPGA. In future versions, SPIDER will include a logic block for extraction, packet building, encoding and serialization at 2.56 Gbits/s. Each channel will then house its own serial link and send the data in streaming mode. The event frame will include all 32 samples in “full readout” mode and typically 8 samples wisely chosen over the signal shape in “smart readout” mode. 
Capture, conversion and extraction occur in parallel in the 8 banks, allowing to process a high rate of events theoretically reaching 50\% of occupancy, including multiple consecutive hits.

\begin{figure}[htbp]
\centering
\includegraphics[width=1\textwidth]{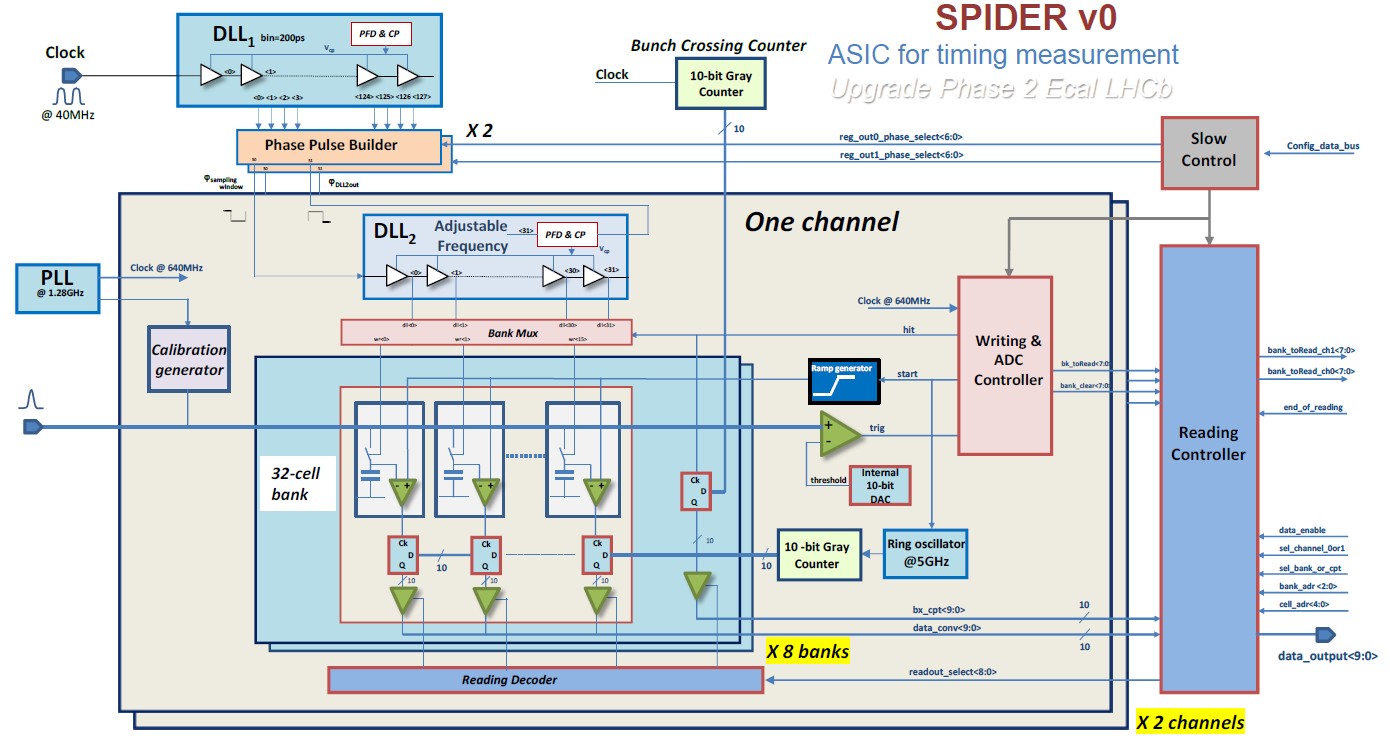}
\caption{SPIDER\_V0 functional block diagram.\label{fig:3}}
\end{figure}

Fig~\ref{fig:3} shows the chip architecture. The control of time windows is based on a master DLL referred to as “DLL1”, locked on the LHC clock, housing 128 delay cells with a step of 195 ps. The “phase pulse builder” then generates, for each channel, pointers based on selected taps of the DLL1. Two of these signals are sent to a second DLL (DLL2) located inside each channel block, one as input of the delay chain and the other as reference for the phase comparator. The DLL2 then generates the 32 write pointers sent to memory cells of the active bank. 
The phase of those 2 signals also defines the beginning and end of the sampling window, and the sampling period directly derives from the duration of the latter.
The phase pulse builder also generates a “trigger enable” window for each channel. Only the triggers occurring inside this window are used for event validation, and others are ignored.

In order to improve the linearity, the analog banks make use of bootstrapped memory cells based on MIM devices.
The A/D conversion follows a massively parallel Wilkinson scheme, based on a ramp generator (typical slope and range of 200 ns for 800 mV) and a 10-bit Gray counter driven by a 5 GHz (tunable) ring oscillator. A great care was taken in the design of the current source of the ramp generator, for both noise and linearity purposes, as well as for the simulations of the chip-wide output bus of the Gray counter, aiming at a maximum skew between the lines below 100 ps p-p. 
In each cell, the conversion block includes buffers, a comparator and a set of flip-flops memorizing the value of the counter when the ramp voltage crosses the cell capacitor voltage. The digital tagging part also includes both a peak and a saturation detectors. The former will be used for smart readout and the latter will provide the width of the saturation plateau.
As usual for analog memories, the chip requires a small set of calibrations (ADC, timing), and all therefore necessary generators are embedded, including a PLL also usable for delivering the state machines clock.

The final layout of SPIDER\_V0 is displayed on Fig~\ref{fig:4}.  DLL1 is located between the 2 channels and the PLL can be seen on bottom left. The chip dimensions are 1960 µm x 4805 µm, dominated in length by the analog memory bank. The future 8-channel version should thus be roughly squared.

\section{Test benches and first results}
\begin{figure}[htbp]
\centering
\includegraphics[width=.4\textwidth]{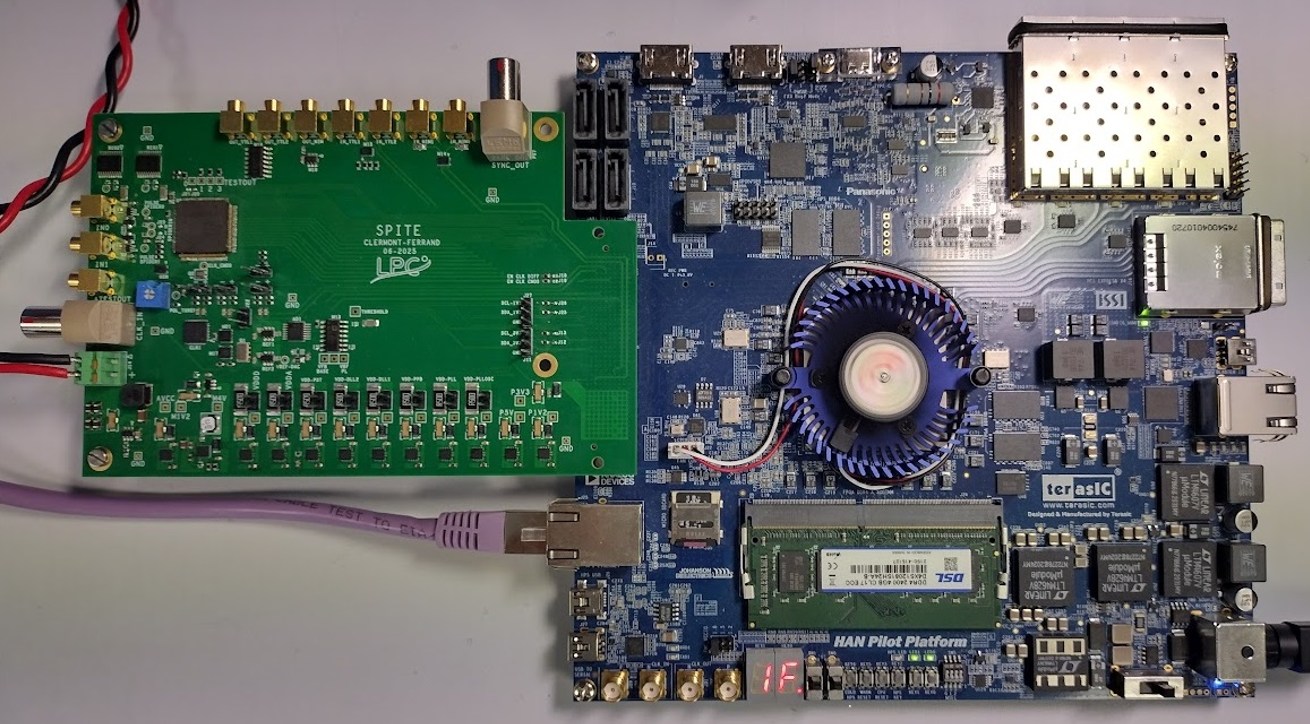}
\quad
\includegraphics[width=.37\textwidth]{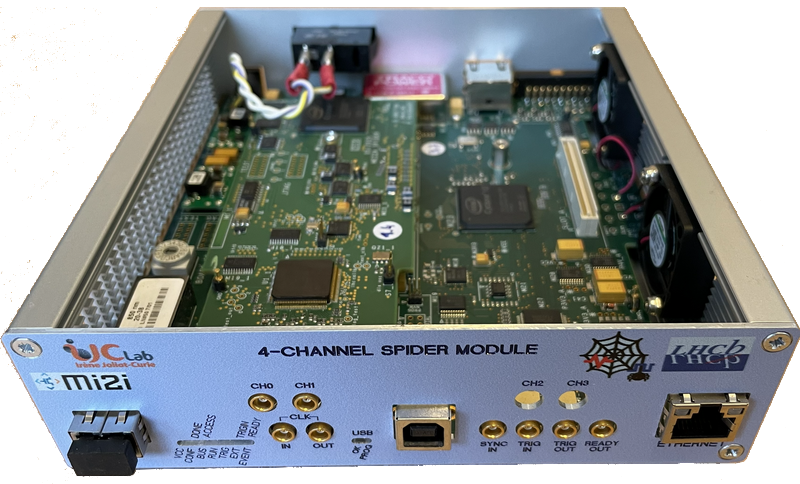}
\quad
\includegraphics[width=0.095\textwidth]{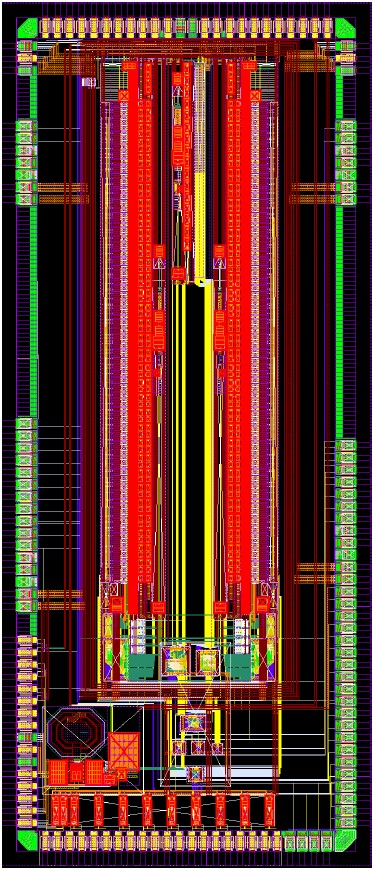}
\caption{Test benches in LPCA (left) and IJCLab (middle). SPIDER\_V0 layout (right).\label{fig:4}}
\end{figure}

Two different test benches have been developed in parallel in IJCLab and LPCA (see Fig:~\ref{fig:4}).
LPCA test bench is based on a custom mezzanine (called “SPITE”) housing SPIDER, mounted on a commercial TerasIC kit.
IJCLab test bench is based on a custom motherboard housing all interfaces equipped with one or two mezzanines (called “Mezza\_Spider”), using the same hardware principle as that of homemade SAMPIC modules. It is integrated into a desktop module and runs with a real time graphical software, which will permit using it soon with 4 channels in beam tests.

The chip was taped out in February 2025. The ASIC dies arrived end of June and got packaged in CQFP128.
Tests started end of July. After a few weeks of debugging, we validated all the main functionalities.
The two DLLs lock with their nominal setup and properly deliver all the pointers to the rest of the chip. The PLL works well but unfortunately doesn't lock at the right frequency. This did not limit the tests since we also have an input for the state machine clock, and it will be easily corrected in the next version. We found a few bugs in the digital part but nothing preventing the chip characterization. The main problem was at the level of the layout where a large crosstalk occurs between the readout data bus and the A/D ramp generator. For the main tests, we thus prevented by firmware having simultaneous conversion and readout. 

We present here the preliminary results. With both channels running, the total power consumption of the digital parts (PLL included) is 64 mW, and as low as 4 mW for the analog parts.

\begin{figure}[htbp]
\centering
\includegraphics[width=1\textwidth]{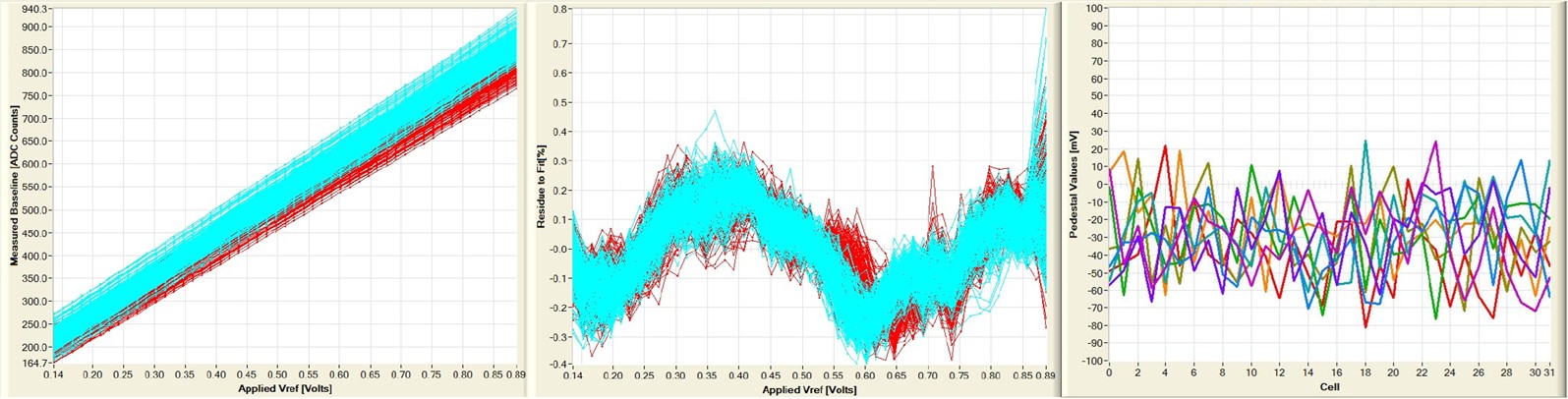}
\caption{ADC calibration (left). Residuals after linear fit in \% (middle) and intercept in mV (right).\label{fig:5}}
\end{figure}

In order to properly visualize signal waveforms, the first calibration to perform concerns the ADC. It measures the dynamic range, the linearity and the fixed pedestal residuals. Results are presented in Fig:~\ref{fig:5}. Dynamic range is of 850 mV, above the target of 800 mV. Linearity is of the order of +-0.2\%, well adapted for timing. Pedestal residuals are rather high (100 mV p-p), but stable over the DC range. We have to further study their precise origin in order to reduce them. 

\begin{figure}[htbp]
\centering
\includegraphics[width=.33\textwidth]{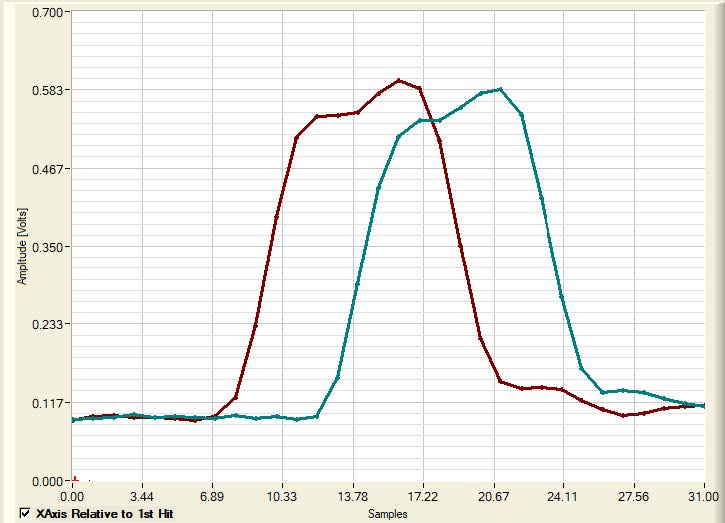}
\includegraphics[width=.33\textwidth]{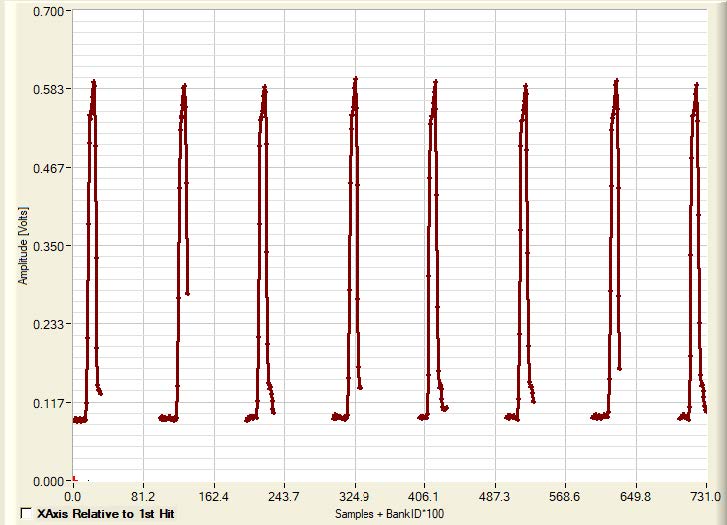}
\includegraphics[width=.31\textwidth]{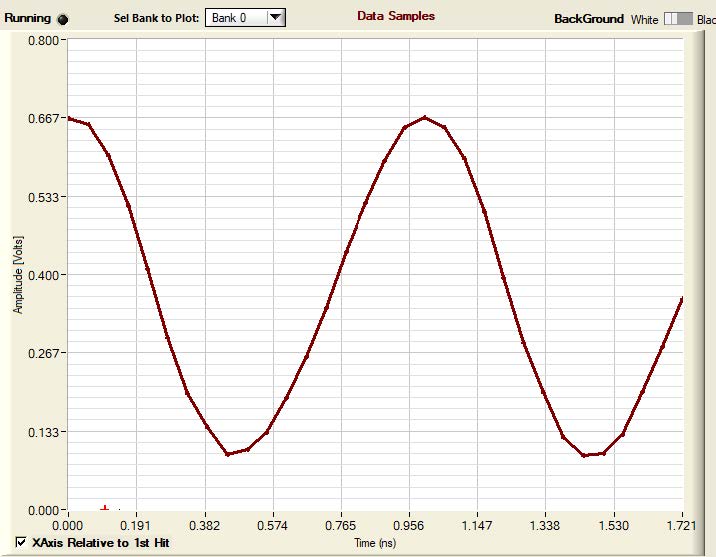}
\caption{2 pulses on 2 channels (left), 1 pulse on the 8 banks of one channel (center), both sampled at 5 GS/s; 1 GHz sinewave sampled at 18 GS/s (right).\label{fig:6}}
\end{figure}

Examples of waveforms are displayed on Fig~\ref{fig:6}. All signals are asynchronous to SPIDER main clock and selected via the trigger enable window. The pulse has an amplitude of 500 mV, a FWHM of 2 ns and a rise time of 800 ps. The middle plot shows the nice functioning of the multi-bank readout. On the right, a 650 mV - 1 GHz sinewave sampled at 18 GS/s (tested up to 25 GS/s).

\begin{figure}[htbp]
\centering
\includegraphics[width=.57\textwidth]{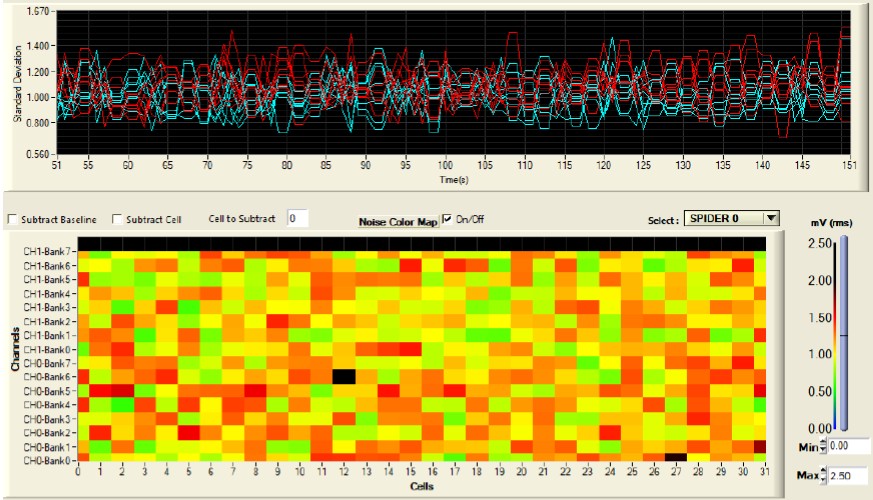}
\quad
\includegraphics[width=.39\textwidth]{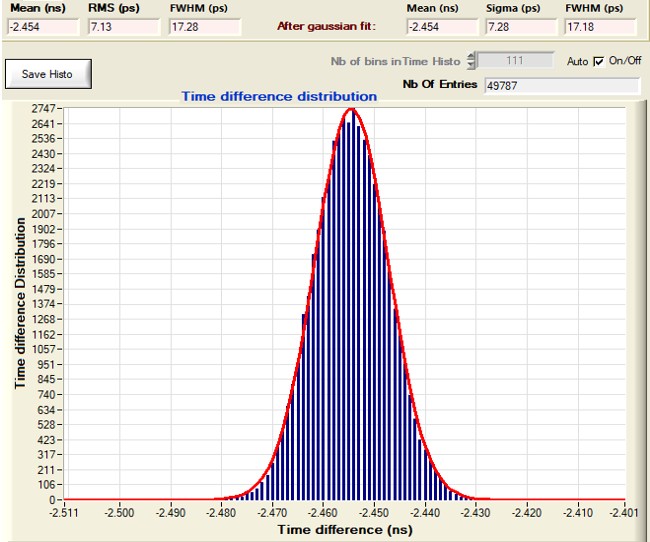}
\caption{Noise performance in mV rms (left) and time resolution on 2 pulses separated by 2.5 ns (right).\label{fig:7}}
\end{figure}

The noise level after ADC calibration is of the order of 1 mV rms (see Fig:~\ref{fig:7} left). As shown on the 3D map, it is homogeneous over all banks and channels. It actually seems to be dominated by the noise on the ADC ramp, which we will further study and try to improve.

The main goal of SPIDER is time measurement. The first tests performed with fully asynchronous pulses and without any time calibration are very encouraging, showing results below 10 ps rms. Indeed, as shown on right of Fig:~\ref{fig:7}, using the 2 pulses shown on left of Fig:~\ref{fig:6} separated by 2.5 ns and sampled at 5 GS/s, we get a TDR (Time Difference Resolution) of 7.13 ps rms, which corresponds to a single channel resolution of 5 ps. With a distance reduced close to zero, future conditions of use in PicoCal, the TDR goes below 6 ps rms. That said, the 2 channels are located inside the same chip and we still have to perform the measurement between 2 different chips.

Next steps are a thorough characterization of this first prototype, followed by a new tape out of the same corrected version (V0+). The 8-channel version is already studied in parallel and will be launched once the V0+ has been validated. The choice of 8 channels is adequate for PicoCal and will permit limiting the length of the timing-critical lines. The results of the V1 will show if increasing the number of channels to 16 could be envisaged.

SPIDER actually targets not only PicoCal, but all fast detectors mounted on current and future accelerators. Its sampling frequency can indeed be adjusted to cope with various signal risetimes. 




\end{document}